\documentclass[submission, Phys]{SciPost}
\hypersetup{
    colorlinks,
    linkcolor={red!50!black},
    citecolor={blue!50!black},
    urlcolor={blue!80!black}
}

\usepackage{graphicx}
\usepackage{amsmath,amssymb}
\usepackage{bm}
\usepackage{braket}
\usepackage{dcolumn}% Align table columns on decimal point
\usepackage[bitstream-charter]{mathdesign}
\DeclareSymbolFont{usualmathcal}{OMS}{cmsy}{m}{n}
\DeclareSymbolFontAlphabet{\mathcal}{usualmathcal}

\newcommand{\diff}[2]{\frac{d#1}{d#2}}

\newcommand{\fdiff}[2]{\frac{\delta #1}{\delta #2}}
\newcommand{\new}{\nonumber\\}
\newcommand{\bx}{\bm{x}}
\newcommand{\bq}{\bm{q}}

\newcommand{\bv}{\bm{v}}

\newcommand{\be}{\bm{e}}

\newcommand{\abs}[1]{\left|#1\right|}
\newcommand{\ave}[1]{\left\langle #1\right\rangle}

\begin{document}
\begin{center}{\Large \textbf{
Dynamical renormalization group analysis of $O(n)$ model in steady shear flow\\
%Scaling theory of $O(n)$ model under shear\\
 }}
\end{center}

% TODO: write the author list here. Use first name (+ other initials) + surname format.
% Separate subsequent authors by a comma, omit comma and use "and" for the last author.
% Mark the corresponding author with a superscript star.
\begin{center}
%Dee E. Faa\textsuperscript{1},
%Aah B. Cee\textsuperscript{2} and
Harukuni Ikeda\textsuperscript{1$\star$},
 and Hiroyoshi Nakano\textsuperscript{2}
\end{center}

\begin{center}
% {\bf 1} Affiliation1
% \\
% {\bf 2} Affiliation2
% \\
{\bf 1} Yukawa Institute for Theoretical Physics, Kyoto University,
Kyoto 606-8502, Japan
\\
{\bf 2} Institute for Solid State Physics, University of Tokyo, 5-1-5,
Kashiwanoha, Kashiwa 277-8581, Japan
\\
${}^\star$ {\small \sf harukuni.ikeda@yukawa.kyoto-u.ac.jp}
\end{center}

\begin{center}
\today
\end{center}

% For convenience during refereeing (optional),
% you can turn on line numbers by uncommenting the next line:
%\linenumbers

\section*{Abstract} 
{\bf
% TODO: write your abstract here.
We study the critical behavior of the $O(n)$ model under steady shear
flow using a dynamical renormalization group (RG) method.  Incorporating
the strong anisotropy in scaling ansatz, which has been neglected in
earlier RG analyses, we identify a new stable Gaussian fixed point.
This fixed point reproduces the anisotropic scaling of static and
dynamical critical exponents for both non-conserved (Model A) and
conserved (Model B) order parameters.  Notably, the upper critical
dimensions are $d_{\text{up}} = 2$ for the non-conserved order parameter
(Model A) and $d_{\text{up}} = 0$ for the conserved order parameter
(Model B), implying that the mean-field critical exponents are observed
even in both $d=2$ and $3$ dimensions.  Furthermore, the scaling
exponent of the order parameter is negative for all dimensions $d \geq
2$, indicating that shear flow stabilizes the long-range order
associated with continuous symmetry breaking even in $d = 2$.  In other
words, the lower critical dimensions are $d_{\rm low} < 2$ for both
types of order parameters.  This contrasts with equilibrium systems,
where the Hohenberg---Mermin---Wagner theorem prohibits continuous
symmetry breaking in $d = 2$.  }

\vspace{10pt}
\noindent\rule{\textwidth}{1pt}
\tableofcontents\thispagestyle{fancy}
\noindent\rule{\textwidth}{1pt}
\vspace{10pt}

\section{Introduction}

The $O(n)$ model is a cornerstone for studying critical phenomena and
encompasses a wide range of phase
transitions~\cite{nishimori2011elements,altland2010condensed}.  For
$n=1$, it describes transitions with discrete $Z_2$ symmetry breaking,
such as Ising magnets and the liquid-gas transition.  For $n\geq 2$, it
captures continuous symmetry breaking, as observed in superfluid $^4$He
and Heisenberg magnets.  Moreover, this model provides insights into the
behavior of liquid crystals.  In equilibrium, the critical behavior of
the $O(n)$ model is now well understood, thanks to advanced techniques
in statistical mechanics, including renormalization group (RG) methods,
exact solutions, and extensive numerical
simulations~\cite{nishimori2011elements,altland2010condensed}.  However,
far from equilibrium, our understanding of the $O(n)$ model is still
under construction.

One common way to drive a system out of equilibrium is by applying
external driving forces, such as shear~\cite{giomi2022ax, giomi2022ex,
krommydas2023tk}.  In 1976, P.~G.~De~Gennes theoretically investigated
the scaling of the non-conserved order parameter near the second-order
phase transition point in the steady shear flow~\cite{de1976effect}.
His mean-field analysis predicts that the shear flow suppresses the
critical fluctuations along the flow direction.  As a consequence, at
the critical point, the correlation function in the Fourier space for
small wave vector $\bq$ exhibits the scaling $C(\bq)\sim
\abs{q_1}^{-2/3}$, where $q_1$ denotes the wave vector along the shear
flow, instead of the standard Ornstein-Zernike like behavior $C(\bq)\sim
\abs{\bq}^{-2}$.  Also, the dynamical scaling argument predicts that the
correlation length $\xi$ and the relaxation time $\tau$ satisfy the
scaling relation $\tau\sim \xi^{3/2}$~\cite{de1976effect}.  This scaling
has been confirmed experimentally in nematic to smectic phase
transition~\cite{safinya1991}.

Subsequently, in 1979, Onuki and Kawasaki investigated the critical
behavior of a conserved order parameter coupled to hydrodynamic
variables in the presence of shear flow (Model H).  Using a dynamical
renormalization group (RG) approach, they predicted that the static
critical exponents in three dimensions are described by the mean-field
theory.  This prediction was later confirmed in experiments on binary
fluids in $d=3$~\cite{beysens1980,beysens1983}.  Given that the upper
critical dimension $d_{\rm up}$ in equilibrium is $d_{\rm up}=4$, this
result highlights a key distinction between equilibrium and
nonequilibrium systems: shear flow can reduce the upper critical
dimension.

The theoretical work of Onuki and Kawasaki has motivated
numerous numerical investigations using the two-dimensional sheared
Ising model.  While the scaling $C(\bq)\sim \abs{q_1}^{-2/3}$ has been
explicitly confirmed by Monte Carlo simulations (Winter et
al.~\cite{winter2010}, Angst et al.~\cite{angst2012}), the value of
critical exponent $\beta$ remains elusive.  The simulations of the
non-conserved sheared Ising model consistently yielded values below
$1/2$, ranging from $\beta = 0.37$ (Winter et al.~\cite{winter2010}) to
$\beta = 0.39 \pm 0.01$ (Saracco and Gonnella~\cite{saracco2009}).
However, the simulation of the conserved model by Saracco and
Gonnella~\cite{saracco2021critical} showed a wider range of values for
$\beta=0.33-0.60$, some of which are close to $1/2$.  Interestingly, the
simulation of the two-dimensional sheared $O(2)$ model by Nakano et
al.~\cite{nakano2021} obtained $\beta=0.48$, a value much closer to the
mean-field prediction.  These conflicting results necessitate further
investigation to resolve the controversy surrounding the mean-field
characters in the sheared models.  

More recently, the numerical simulation of the sheared $O(2)$
model~\cite{nakano2021} revealed the occurrence of long-range order
associated with continuous symmetry breaking even in $d=2$ dimension.
For the $O(2)$ model, the order parameter fluctuations can be decomposed
into phase and amplitude fluctuations. The phase fluctuations, which are
soft modes, are referred to as Nambu-Goldstone (NG)
modes~\cite{minami2021,nakano2021}.  It was found that shear flow not
only modifies the critical fluctuations but also alters the fluctuations
of the NG modes from the standard Ornstein-Zernike-like behavior,
$C(\mathbf{q})\sim \abs{\mathbf{q}}^{-2}$, to a fractional scaling,
$C(q_1)\sim \abs{q_1}^{-2/3}$, thereby stabilizing the long-range order
in two dimensions.  This two-dimensional continuous symmetry breaking is
remarkable because, in equilibrium systems, the lower critical
dimension, $d_{\rm low}$, is $2$, and the Hohenberg-Mermin-Wagner
theorem prohibits continuous symmetry breaking in two
dimensions~\cite{hohenberg1967,mermin1966}.  This reduction of the lower
critical dimension further highlights a key distinction between
equilibrium and nonequilibrium systems.

The reduction of $d_{\rm up}$ and $d_{\rm low}$ due to shear flow is a
significant discovery in the study of non-equilibrium phase transitions.
However, a complete theoretical understanding of this phenomenon is
still lacking.  The dynamical RG method offers a powerful approach to
tackle this challenge.
% Although Onuki and Kawasaki's work in 1979
% (OK79)~\cite{onuki1979} previously applied this method for Model H,
% their analysis did not incorporate the anisotropy inherent to sheared
% systems, and the resulting fixed point was unstable under shear. 
The aim of this paper is to propose that a stable fixed point can be
obtained by correctly accounting for the anisotropy of the sheared
system.  The dynamical RG methods for anisotropic systems have been
developed in several non-equilibrium systems, such as the directed
percolation~\cite{broadbent1957,takeuchi2007,takeuchi2009,henkel2008non},
growing
interfaces~\cite{kardar1986,burger1989,hwa1989,hwa1992,bray2001}, polar
flocks~\cite{vicsek1995,toner1995,toner1998,toner2012,mahault2019,ikeda2024minimum,chate2024},
and coarsening dynamics subjected to the external
field~\cite{janssen1986ra, leung1986ax} and the shear
flow~\cite{bray2000,bray2001in,corberi2002,corberi2003}.  In this work,
we apply the anisotropic dynamical RG formalism developed in these
studies to the $O(n)$ model in the steady shear flow.  We show that the
anisotropic scaling ansatz leads to a new Gaussian fixed point, which is
stable against shear.  The upper critical dimension of the new fixed
point is $d_{\rm up}=2$ for a non-conserved order parameter (Model A)
and $d_{\rm up}=0$ for a conserved order parameter (Model B), meaning
that the mean-field critical exponents are observed in $d=2$ and $3$.

The remainder of this paper is organized as follows.  In
Sec.~\ref{170051_7Aug24}, we investigate the model for the
non-conservative order parameter (Model A).  In
Sec.~\ref{203247_28Jul24}, we investigate Model B in the steady shear.
In Sec.~\ref{203110_28Jul24}, we conclude the work.

\section{Model A}
\label{170051_7Aug24}

\subsection{Settings}
\label{192515_28Jul24}

\begin{figure}[t]
\begin{center}
\includegraphics[width=10cm]{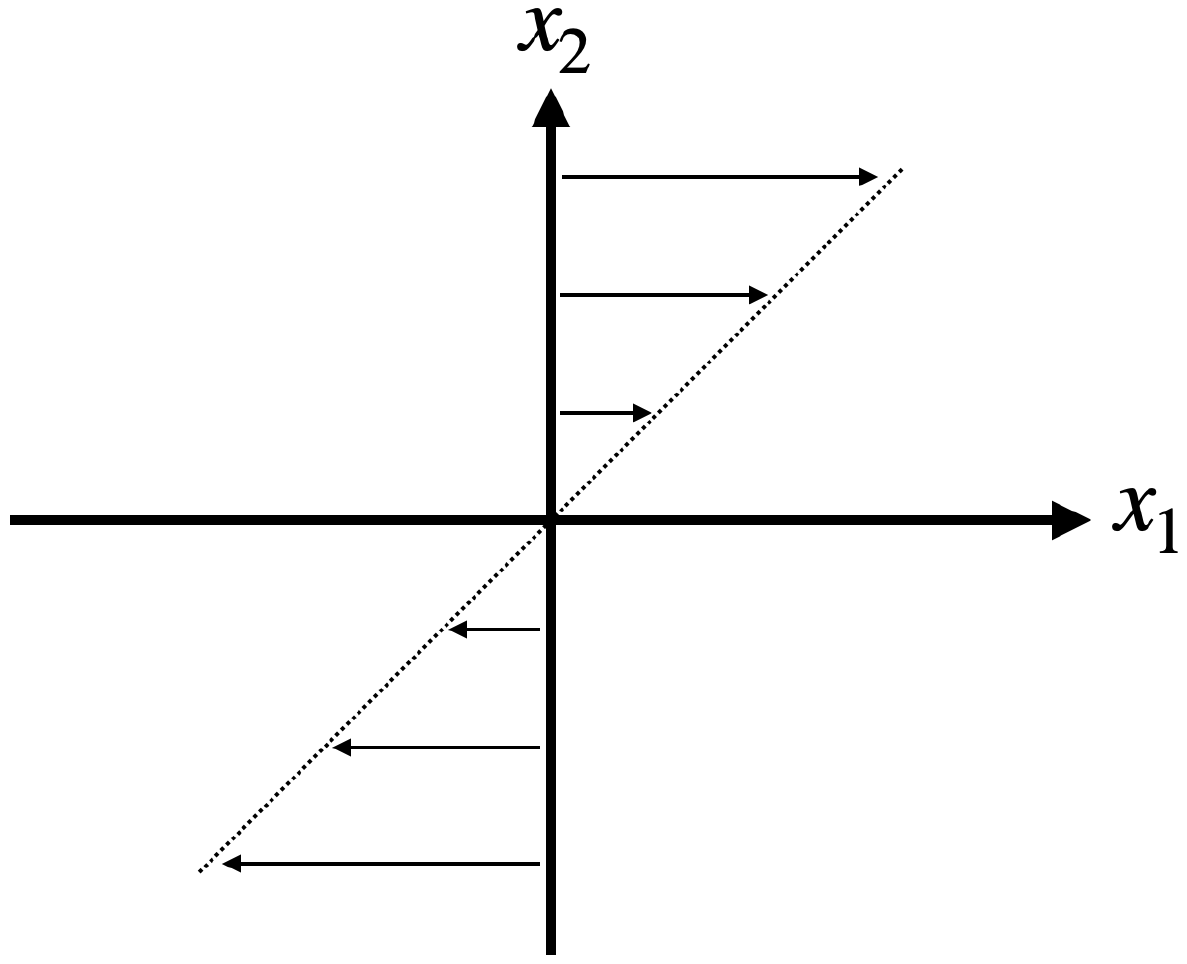} \caption{
Schematic picture of the shear flow. }
				       \label{211626_28Jul24}
\end{center}
\end{figure}

We consider the $d$-dimensional $O(n)$ model subjected to the steady
shear~\cite{minami2021,nakano2021,minami2022origin}:
\begin{align}
\dot{\phi}_a+\bm{v}\cdot\nabla\phi_a
= D\nabla^2\phi_a
- \fdiff{F[\vec{\phi}]}{\phi_a}
+\sqrt{2\Delta}\xi_a,
\label{201609_30Dec23}
\end{align}
where the $n$-component vector $\vec{\phi}(\bx,t)=\{\phi_1(\bx,t),\cdots, \phi_n(\bx,t)\}$
denotes the order parameter at position $\bx=\{x_1,\cdots, x_d\}$ and time $t$, 
\begin{align}
F[\vec{\phi}] =  \int d\bx \left[\frac{\varepsilon}{2}\left(\vec{\phi}
\cdot\vec{\phi}\right)
+ \frac{u}{4}\left(\vec{\phi}\cdot\vec{\phi}\right)^2\right]\label{062606_1Jan24}
\end{align}
denotes the standard $\phi^4$ free-energy, 
and $\xi_a(\bx,t)$ denotes the white noise whose mean and variance are given by
\begin{align}
&\ave{\xi_{a}(\bx,t)} = 0,\new
&\ave{\xi_{a}(\bx,t)\xi_{b}(\bx',t')}
 = \delta_{ab}\delta(\bx-\bx')\delta(t-t').
\label{021726_31Dec23}
\end{align}
The linear advection term in Eq.~(\ref{201609_30Dec23}),
$\bv\cdot\nabla\phi_a$, represents the effect of the shear flow.  We
consider the simple shear flow along the $x_1$ axis with a constant
gradient along the $x_2$ axis:
\begin{align}
\bm{v} = \{\dot{\gamma}x_2,0,\cdots, 0\},
\label{024236_31Dec23}
\end{align}
where $\dot{\gamma}$ denotes the shear rate, see Fig.~\ref{211626_28Jul24}.

The $O(n)$ model for $n=1$ describes the Ising universality class.  In
previous works, the Ising model in shear has been studied
extensively~\cite{cirillo2005,saracco2009,hucht2009,winter2010,angst2012}.
In particular, for the infinitely large shear rate
$\dot{\gamma}\to\infty$, the mean-field approximation becomes exact,
which enables to solve the model analytically~\cite{hucht2009}.
However, numerical simulations have produced the conflicting results, as
summarized in
Introduction~\cite{cirillo2005,saracco2009,hucht2009,winter2010,angst2012}.
The $O(n)$ model in shear with $n=2$ also has been studied
extensively~\cite{nakano2021,minami2021,minami2022origin}.  A recent
numerical simulation has demonstrated that the model undergoes the
continuous symmetry breaking even in $d=2$.  Interestingly, the critical
exponents of the transition agree with those of the mean-field
prediction~\cite{minami2021,minami2022origin}.

\subsection{Renormalization Group flow equations}
\label{192536_28Jul24}

\begin{figure}[t]
\begin{center}
\includegraphics[width=10cm]{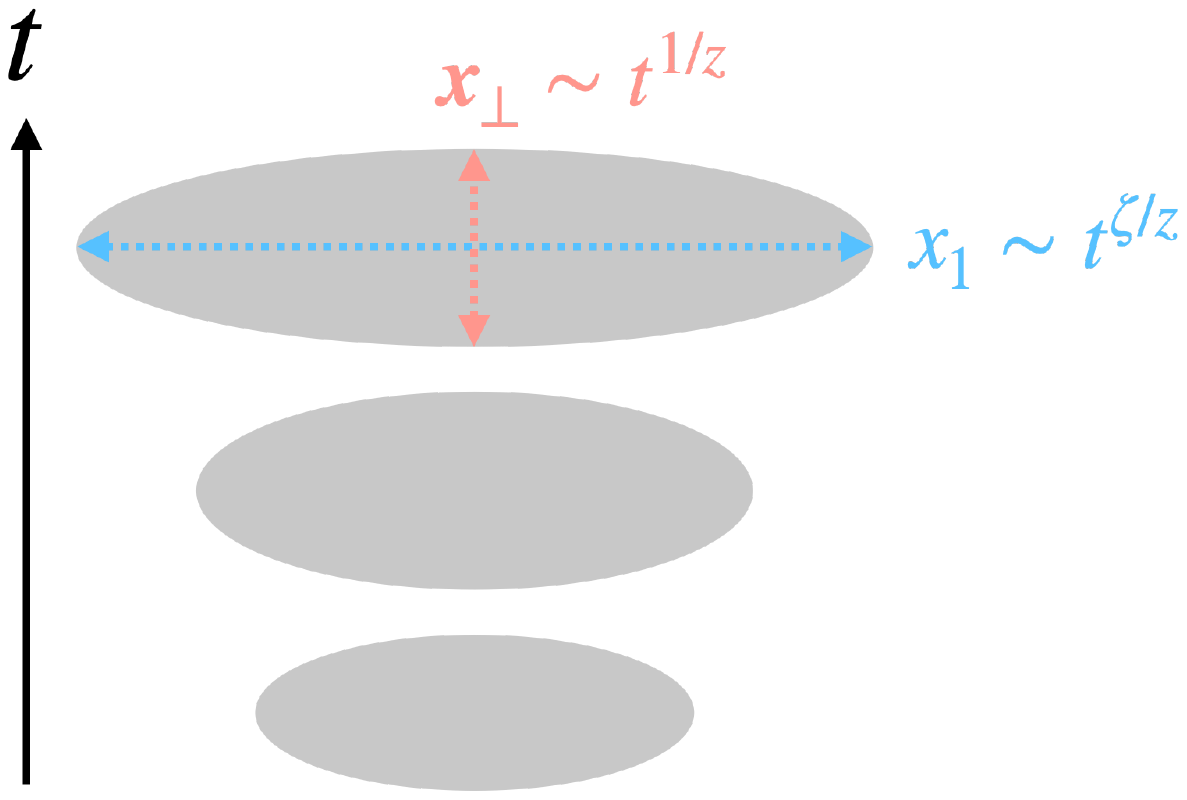} \caption{ Schematic picture of
the typical size of the critical fluctuation.}  \label{135255_4Aug24}
\end{center}
\end{figure}

To investigate the large spatiotemporal behavior of the model near the
critical point $\varepsilon=0$, we consider the anisotropic scaling
transformations~\cite{hwa1989,toner1995,bray2001in}:
\begin{align}
x_1= l^\zeta x_1',\
\bx_\perp = l\bx_\perp',\
t= l^z t',\
\phi_a= l^\chi \phi_a',\label{002440_14Jul24}
\end{align}
where $x_1$ and $\bx_\perp=\{x_2,\cdots, x_d\}$ respectively denote the
coordinates parallel and perpendicular to the flow direction. The above
scaling is tantamount to assuming that the typical size of the
fluctuations grows anisotropically as $x_1\sim t^{\zeta/z}$ and
$\bx_\perp \sim t^{1/z}$, see Fig.~\ref{135255_4Aug24}.  Substituting
the scaling relations~(\ref{002440_14Jul24}) into the equation of motion
(\ref{201609_30Dec23}) and dividing both sides by $l^{\chi-z}$, we get
\begin{align}
&\dot{\phi}_a' + l^{z+1-\zeta}\dot{\gamma}x_2'\partial_{1}' \phi_a'
 \new
 &=l^{z-2\zeta}D_{\parallel}(\partial_{1}')^2 \phi_a' +
 l^{z-2}D_\perp (\nabla_\perp')^2 \phi_a'
-l^z \varepsilon\phi_a' -l^{z+2\chi}u(\vec{\phi}'\cdot\vec{\phi}')\phi_a' 
+ l^{\frac{z-2\chi-(d-1+\zeta)}{2}}
 \sqrt{2\Delta}\xi_a',
\end{align}
where $D_\parallel = D_\perp = D$. The above equation implies that
under the scaling transformations~(\ref{002440_14Jul24}), 
the
coefficients of the equation of motion~(\ref{201609_30Dec23}) 
are transformed to 
\begin{align}
&\dot{\gamma}' = l^{z+1-\zeta}\dot{\gamma},\new 
&D_{\parallel}' = l^{z-2\zeta}D_{\parallel},\new 
&D_{\perp}' = l^{z-2}D_{\perp},\new 
&\varepsilon' = l^z \varepsilon,\new 
&u' = l^{z+2\chi}u,\new 
&\Delta' = l^{z-2\chi-(d-1+\zeta)}\Delta,\label{002331_14Jul24}
\end{align}
leading to the RG flow equations:
\begin{align}
\left.\diff{\dot{\gamma}}{l}\right|_{l=1} &= 
(z+1-\zeta)\dot{\gamma},\label{140102_13Jul24}\\
\left.\diff{D_{\parallel}}{l}\right|_{l=1} &= 
(z-2\zeta)D_{\parallel},\label{164721_14Jul24}\\
\left.\diff{D_{\perp}}{l}\right|_{l=1} &= (z-2)D_{\perp},\label{164732_14Jul24}\\
\left.\diff{\Delta}{l}\right|_{l=1} &
 = \left[z-2\chi-(d-1+\zeta)\right]\Delta,\label{164749_14Jul24}\\
\left.\diff{\varepsilon}{l}\right|_{l=1} &= z\varepsilon,\\
\left.\diff{u}{l}\right|_{l=1} &= (z+2\chi)u.\label{120212_14Jul24}
\end{align}
The equations are derived by a naive scaling argument, which is
tantamount to neglecting the graphical
corrections~\cite{ikeda2024ar}. The equations can be applied only near
the Gaussian fixed point.  The stability of the Gaussian fixed point can
be discussed by the RG flow equation of the non-linear term,
Eq.~(\ref{120212_14Jul24}), as we will discuss below.

\subsection{Gaussian fixed point without shear}
\label{192600_28Jul24}

Before investigating the critical behavior under shear flow, we review
the RG analysis for the equilibrium case ($\dot{\gamma}=0$) of a
non-conserved order parameter (Model A) to establish a baseline for
comparison.  For $\dot{\gamma}=0$, Eq.~(\ref{140102_13Jul24}) is
automatically satisfied.  We assume that the transport coefficients and
the noise strength are scale-invariant, which requires
\begin{align}
\left.\diff{D_{\parallel}}{l}\right|_{l=1}
=\left.\diff{D_{\perp}}{l}\right|_{l=1}
=\left.\diff{\Delta}{l}\right|_{l=1}=0,\label{115404_4Aug24}
\end{align}
leading to the scaling relations
\begin{align}
z-2\zeta = z-2 = z-2\chi-(d-1+\zeta) = 0.\label{115653_4Aug24}
\end{align}
This can be solved easily, and we get 
\begin{align}
\zeta = 1,\quad z = 2,\quad \chi = \frac{2-d}{2}.\label{115738_14Jul24} 
\end{align}
The results are consistent with those of the Gaussian fixed point in
equilibrium~\cite{nishimori2011elements}.

First, we consider the upper critical dimension, $d_{\rm up}$.  Above
this dimension, the critical behavior of the system is captured by
mean-field theory, resulting in mean-field critical exponents.  To
determine the upper critical dimensions, $d_{\rm up}$, we analyze the
stability of the Gaussian fixed point against the non-linear term $u$.
By substituting the exponents~(\ref{115738_14Jul24}) into the RG flow
equation for the non-linear term~(\ref{120212_14Jul24}), we have
\begin{align}
 \left.\diff{u}{l}\right|_{l=1} = (4-d)u,
\end{align}
leading to $u\sim l^{4-d}$. For $d<4$, the coefficient diverges
$u\to\infty$ in the thermodynamic limit $l\to\infty$, which destabilizes
the Gaussian fixed point. On the contrary, for $d>4$, the coefficient
vanishes $u\to 0$, meaning that the Gaussian fixed point is
stable. Therefore, the upper critical dimension $d_{\rm up}$ is
\begin{align}
 d_{\rm up} = 4.
\end{align}

Note that $u$ is a dangerously irrelevant variable above the upper
critical dimensions $d>d_{\rm up}$~\cite{nishimori2011elements}.  In
other words, $u$ is irrelevant but still affects the scaling.  To
understand this, let us write the scaling law of $\phi_a$ in the ordered
phase $\varepsilon<0$ with the variable $u$ explicitly included as
\begin{align}
\phi_a(\varepsilon,u) = l^{\chi}
\phi_a (l^z \varepsilon, l^{z+2\chi}u).
\end{align}
Setting $l=\abs{\varepsilon}^{-1/z}$, 
we get 
\begin{align}
\phi_a(\varepsilon,u) = \abs{\varepsilon}^{-\chi/z}
\phi_a (-1, \abs{\varepsilon}^{-(z+2\chi)/z}u).\label{190931_14Jul24}
\end{align}
One may naively expect that $\phi_a (-1, u)$ is analytic in $u$ and
$\phi_a (-1, u) \sim u^0$ for $u\ll 1$, leading to $\ave{\phi_a}\sim
\abs{\varepsilon}^{\beta}$ with $\beta=-\chi/z=d/4$.  However, in fact,
$\phi_a(-1, u)$ is not analytic in $u$ for $d>d_{\rm up}$.  This can be
seen from the saddle point equation of the free-energy
(\ref{062606_1Jan24}),
\begin{align}
\fdiff{F}{\phi_a} = \varepsilon\phi_a + u(\vec{\phi}\cdot\vec{\phi})\phi_a = 0, 
\end{align}
which implies $\phi_a(-1,u)\sim u^{-1/2}$ for $u\ll 1$.  Consequently,
we have $\phi_a (-1, u) \sim u^{-1/2}$ instead of $\phi_a (-1, u) \sim
u^0$, and thus
\begin{align}
\phi_a(\varepsilon,u) \sim \abs{\varepsilon}^{-\chi/z}
\left(\abs{\varepsilon}^{-(z+2\chi)/z}u\right)^{-1/2}\sim \abs{\varepsilon}^{1/2}.
\end{align}
for $\abs{\varepsilon}\ll 1$.
Therefore, the correct critical exponent above $d_{\rm up}$ is 
\begin{align}
\beta = \frac{1}{2}.\label{123953_16Jul24}
\end{align}
For more detailed discussions, see, for instance, Sec.~4.~2 in
Ref.~\cite{nishimori2011elements}.

Next, we study the lower critical dimension $d_{\rm low}$.  It is
defined as the dimension below which the long-range order associated
with continuous symmetry breaking is destroyed by fluctuations.  For the
$O(n)$ model for $n\geq 2$, consider the ordered phase ($\varepsilon<0$)
where the order parameter is oriented along the $a=1$ direction, such as
$\ave{\phi_a}=(-\varepsilon/u)^{1/2}\delta_{a1}$.  The fluctuations of
the order parameter perpendicular to this direction, denoted by $\delta
\phi_a$ ($a\neq 1$), correspond to the NG
modes~\cite{nambu1960,goldstone1962}.  The NG modes become divergent
below the lower critical dimension $d_{\rm low}$, destroying the
long-range order. To calculate $d_{\rm low}$, we observe the scaling for
$\delta \phi_a$ ($a\neq 1$)~\cite{toner1995}.  They follow the dynamics
$\delta\dot{\phi}_a=D\nabla^2\delta\phi_a + \sqrt{2\Delta}\xi_a$ for
$a\neq 1$. A similar scaling analysis as above leads to
\begin{align}
\ave{\delta \phi_a^2}\sim l^{2\chi}.
\end{align}
If $\chi>0$, the fluctuations diverge in the thermodynamic limit $l\to
\infty$, meaning that a necessary condition for the continuous symmetry
breaking is $\chi=(2-d)/2<0$, or equivalently, $d>2$. Therefore, the
lower critical dimension is
\begin{align}
d_{\rm low}=2,
\end{align}
which is consistent with the Hohenberg--Mermin--Wagner
theorem~\cite{hohenberg1967,mermin1966}.

Finally, we discuss that the Gaussian fixed point is destabilized by the
shear in any $d$. Substituting (\ref{115738_14Jul24}) into the RG flow
equation of $\dot{\gamma}$~(\ref{140102_13Jul24}), we get
\begin{align}
 \left.\diff{\dot{\gamma}}{l}\right|_{l=1} = 2\dot{\gamma},
\end{align}
leading to $\dot{\gamma}\sim l^2$. This means that the shear rate
$\dot{\gamma}$ is a relevant variable, which destabilizes the
equilibrium Gaussian fixed point in any spatial dimension
$d$. Therefore, we need to seek a new stable fixed point, which will be
discussed in the next section.

\subsection{Gaussian fixed point with shear}
\label{193406_28Jul24} 

Now, we investigate the model with the finite shear rate
$\dot{\gamma}\neq 0$. For this purpose, we observe the scaling behaviors
of the shear rate $\dot{\gamma}$, the transport coefficients
$D_\parallel$, $D_\perp$, and the strength of the noise $\Delta$, whose
RG flow equations are given by
(\ref{140102_13Jul24})-(\ref{164749_14Jul24}).  To determine the three
independent exponents, $\zeta$, $z$, and $\chi$, we need three
independent conditions. For instance, the scale invariance of
$D_\parallel$, $D_\perp$, and $\Delta$ leads to the equilibrium Gaussian
fixed point, which is unstable against shear, as discussed in the
previous subsection. Another physically meaningful solution can be
obtained by requiring the scale invariance of $\dot{\gamma}$, $D_\perp$,
and $\Delta$:
\begin{align}
\left.\diff{\dot{\gamma}}{l}\right|_{l=1}
=\left.\diff{D_{\perp}}{l}\right|_{l=1}
=\left.\diff{\Delta}{l}\right|_{l=1}=0,
\end{align}
or equivalently, 
\begin{align}
z+1-\zeta = z-2 = z-2\chi-(d-1+\zeta) = 0,
\end{align}
leading to 
\begin{align}
\zeta = 3,\quad z = 2,\quad \chi = -\frac{d}{2}.\label{123941_16Jul24} 
\end{align}
Any other choices yield unphysical results~\footnote{The scale
invariance of $\dot{\gamma}$, $D_\parallel$, and $\Delta$ leads to a
negative value of the dynamical critical exponent $z=-2$, which is
unphysical.  The scale invariance of $\dot{\gamma}$, $D_\parallel$, and
$D_\perp$ leads to $z+1-\zeta=z-2\zeta=z-2=0$, which does not have the
solution.}. The scaling $x_\parallel \sim l^\zeta$ and $t\sim l^z$ imply
the scaling relation between the relaxation time $\tau$ and correlation
length along the flow direction $\xi_\parallel$: $\tau\sim
\xi_\parallel^{2/3}$, which is consistent with previous
studies~\cite{de1976effect,safinya1991,hucht2009}. Substituting the
above exponents into Eq.~(\ref{164721_14Jul24}), we get
\begin{align}
\left.\diff{D_\parallel}{l}\right|_{l=1} =  -4D_\parallel,
\end{align}
leading to $D_\parallel\sim l^{-4}$, which vanishes in the thermodynamic
limit $l\to\infty$. Therefore, the diffusion term parallel to the flow
direction is irrelevant.

In principle, one may also consider higher-order nonlinear terms and
higher-derivative terms. Under the anisotropic scaling around the
Gaussian fixed point adopted here, however, such terms are more
irrelevant than those retained in the present effective theory. For this
reason, they do not modify the renormalization-group analysis at the
order discussed in this paper.

The upper critical dimension $d_{\rm up}$ is calculated by observing the
RG flow equation for the non-linear term (\ref{120212_14Jul24}):
\begin{align}
\left.\diff{u}{l}\right|_{l=1} = (2-d)u,
\end{align}
leading to $u\sim l^{2-d}$. For $d>2$, $u$ vanishes in the
thermodynamic limit $l\to\infty$, meaning that the upper critical
dimension is
\begin{align}
d_{\rm up}= 2,\label{214907_5Aug24}
\end{align}
This suggests that the mean-field critical exponents are observed in
physical dimensions $d=2$ and $3$.  Note that, similar to the
equilibrium case, $u$ remains a dangerously irrelevant variable.
Assuming the same scaling behavior as in equilibrium, namely $\phi_a
(-1, u) \sim u^{-1/2}$ instead of $\phi_a (-1, u) \sim u^0$, we obtain
\begin{align}
\phi_a(\varepsilon,u) \sim \abs{\varepsilon}^{-\chi/z}
\left(\abs{\varepsilon}^{-(z+2\chi)/z}u\right)^{-1/2}\sim \abs{\varepsilon}^{1/2}.
\end{align}
for $\abs{\varepsilon}\ll 1$.
Therefore, the critical exponent $\beta$ above $d_{\rm up}=2$ is 
\begin{align}
\beta = \frac{1}{2},
\end{align}
which is consistent with theoretical studies in the limit of large shear
rate~\cite{hucht2009}. However, as mentioned in the Introduction,
previous numerical studies in two dimensions have often reported values
of $\beta$ that deviate from
$1/2$~\cite{cirillo2005,saracco2009,hucht2009,winter2010,angst2012}.
This discrepancy may be attributed to the presence of logarithmic
corrections at the upper critical
dimension~\cite{nishimori2011elements}.  Future numerical work should
investigate this possibility.

The lower critical dimension $d_{\rm low}$ is also varied in the
presence of shear flow. Indeed, $\chi<0$ for any $d>0$, which means that
$\ave{\delta \phi_a^2}\sim l^{2\chi} \to 0$, and then the lower critical
dimension $d_{\rm low}$ turns out to be
\begin{align}
d_{\rm low}= 0.\label{214933_5Aug24}
\end{align}
In particular, the continuous symmetry breaking can occur even in $d=2$,
which is prohibited in equilibrium by the Hohenberg--Mermin--Wagner
theorem~\cite{hohenberg1967,mermin1966}.  The result is consistent with
a recent numerical simulation for the $O(2)$ model in
$d=2$~\cite{nakano2021}.

\subsection{Correlation functions}
\label{202615_28Jul24} The presence of shear flow significantly alters
the behavior of correlation functions, both for the critical
fluctuations and the NG modes.  We first investigate the correlation
function just above the critical point ($\varepsilon>0$). The scaling
behaviors (\ref{002440_14Jul24}) and (\ref{002331_14Jul24}) lead to
\begin{align}
C(x_1,\bx_\perp,\varepsilon) &
 = \ave{\vec{\phi}(x_1,\bx_\perp)\cdot\vec{\phi}(0,0)}\new 
&=l^{2\chi}C(l^{-\zeta}x_1,l^{-1}\bx_\perp, l^z \varepsilon).
\end{align}
In the Fourier space, we get 
\begin{align}
C(q_1,\bm{q}_\perp,\varepsilon) &
 = \int d\bx e^{i(q_1x_1+\bq_\perp\cdot\bx_\perp)}
C(x_1,\bx_\perp,\varepsilon)\new 
& =l^{2\chi+\zeta+d-1}
C(l^\zeta q_1,l\bm{q}_\perp,l^z\varepsilon)
=  l^{z}
C(l^\zeta q_1,l\bm{q}_\perp,l^z\varepsilon)\label{164347_19Jul24}
\end{align}
Substituting $l=\varepsilon^{-1/z}$, we get
\begin{align}
C(q_1,\bq_\perp,\varepsilon) =\varepsilon^{-1}
C(\varepsilon^{-\nu_\parallel}q_1,\varepsilon^{-\nu_\perp}\bq_\perp,1)
\label{160612_4Aug24}
\end{align}
with 
\begin{align}
&\nu_\parallel = \frac{3}{2},
&\nu_\perp = \frac{1}{2}.\label{100237_5Aug24}
\end{align}
The scaling form (\ref{160612_4Aug24}) implies that the 
correlation lengths parallel and perpendicular to the flow direction,
$\xi_{\parallel}$
and $\xi_{\perp}$,
diverge as 
\begin{align}
&\xi_\parallel \sim \varepsilon^{-\nu_\parallel},
&\xi_\perp \sim \varepsilon^{-\nu_\perp}.
\end{align}
The exponents are consistent with the previous theoretical and numerical
simulations for the Ising-model~\cite{angst2012,hucht2009}, and $O(n)$
model~\cite{nakano2021,minami2021,minami2022origin} in shear.
Repeating the similar scaling analysis, we get 
\begin{align}
&C(q_1,\bm{0},0) \sim \abs{q_1}^{-2/3},&\abs{q_1}\ll 1,\label{ca}\\
&C(0,\bq_\perp,0) \sim \abs{\bq_\perp}^{-2},&\abs{\bq_\perp}\ll 1,\\
&C(0,0,\varepsilon) \sim \varepsilon^{-1},&\varepsilon\ll 1,
\end{align}
implying that the correlation function can be expanded as 
\begin{align}
C(q_1,\bq_\perp,\varepsilon) =
\left[c_1\varepsilon + c_2 \abs{q_1}^{2/3}
 + c_3 \abs{\bq_\perp}^2 + \cdots\right]^{-1},\label{175251_20Jul24} 
\end{align}
where $c_1$, $c_2$, and $c_3$ are some constants. This is consistent
with the previous studies for the Ising model in
shear~\cite{hucht2009,angst2012}. 

We proceed to the analysis of the NG mode.  Following the deviation of
Eq.~(24), we consider the ordered phase ($\varepsilon<0$) where the
order parameter is oriented along the $a=1$ direction, such as
$\ave{\phi_a}=(-\varepsilon/u)^{1/2}\delta_{a1}$.  The NG mode, $\delta
\phi_a$ ($a\neq 1$), is governed by the equation
$\delta\dot{\phi}_a+\dot{\gamma}x_2\partial_{1} \phi_a
=D\nabla^2\delta\phi_a + \sqrt{2\Delta}\xi_a$, and then a similar
scaling analysis as above leads to $C(q_1,\bq_\perp,\varepsilon)\approx
(c_2\abs{q_1}^{2/3}+c_3\abs{\bq_\perp}^2)^{-1}$ for $\varepsilon<0$.
The result is consistent with the recent numerical
simulation~\cite{nakano2021} and linear analysis~\cite{minami2021}.

\section{Model B}
\label{203247_28Jul24}

Model B is a simple variation of Model A, which describes the dynamics
of the conserved order parameter, such as density in the phase
separation~\cite{hohenberg1977}.  In equilibrium, both models share the
same static critical exponents, since their values are determined solely
by the free energy and are independent of the specific dynamics.  Their
dynamical critical exponents differ due to their distinct
dynamics~\cite{hohenberg1977,nishimori2011elements}.  In contrast, we
show that under uniform shear flow, even the static critical exponents
depend on the type of dynamics, leading to different values for Model A
and Model B.  Furthermore, we find that critical fluctuations are more
strongly suppressed in Model B compared to Model A, leading to smaller
values for the critical dimensions, $d_{\rm low}$ and $d_{\rm up}$.

\subsection{Settings}
We consider the following equation of motion for the conserved-order parameter in the steady shear flow (Model B)~\cite{kawasaki1966,hohenberg1977}:
\begin{align}
\dot{\phi}_a+\bm{v}\cdot\nabla\phi_a
=
-\nabla^2\left[
 D\nabla^2\phi_a
+ \fdiff{F[\vec{\phi}]}{\phi_a}\right]
+\sqrt{2\Delta}\nabla\cdot\bm{\xi}_a,
\end{align}
where $F[\vec{\phi}]$ denotes the free-energy~(\ref{062606_1Jan24}),
$\bm{v}$ denotes the velocity of the shear flow~(\ref{024236_31Dec23}),
and ${\bm{\xi}_a=\{\xi_{a,1},\cdots, \xi_{a,d}\}}$ denotes the white
noise whose mean and variance are given by
\begin{align}
&\ave{\xi_{a,i}(\bx,t)} = 0,\new
&\ave{\xi_{a,i}(\bx,t)\xi_{b,j}(\bx',t')}
 = \delta_{ab}\delta_{ij}\delta(\bx-\bx')\delta(t-t').
\end{align}

\subsection{Renormalization Group flow equations}
To investigate the large spatiotemporal behavior of the model, we
consider the anisotropic scaling transformations~(\ref{002440_14Jul24}).
As we will see later, the anisotropic parameter is $\zeta>1$,
which enables the following approximation:
\begin{align}
&\nabla = l^{-\zeta}\partial_1'\be_1 + 
l^{-1}\nabla_\perp' \approx l^{-1}\nabla_\perp',\label{204823_5Aug24}
\end{align}
where $\be_1$ denotes the unit vector along $x_1$.
After some manipulations, we get 
\begin{align}
&\dot{\phi}_a' + l^{z+1-\zeta}\dot{\gamma}x_2'\partial_{1}' \phi_a'\new 
&\approx  l^{z-4}D_\perp (\nabla_\perp')^4 \phi_a'
-l^{z-2} (\nabla_\perp')^2\varepsilon\phi_a'
 -l^{z+2\chi-2}(\nabla_\perp')^2u(\vec{\phi}'\cdot\vec{\phi}')\phi_a' 
  + l^{\frac{z-2\chi-(d-1+\zeta)-2}{2}}
 \sqrt{2\Delta}\nabla_\perp'\cdot\bm{\xi}_a',
\end{align}
leading to the RG flow equations~\cite{nishimori2011elements}:
\begin{align}
\left.\diff{\dot{\gamma}}{l}\right|_{l=1}
 &= (z+1-\zeta)\dot{\gamma},\label{133358_30Aug24}\\
\left.\diff{D_{\perp}}{l}\right|_{l=1}
 &= (z-4)D_{\perp},\\
\left.\diff{\Delta}{l}\right|_{l=1} 
 &= \left[z-2\chi-(d-1+\zeta)-2\right]\Delta,\\ 
\left.\diff{\varepsilon}{l}\right|_{l=1} &= (z-2)\varepsilon,\\
\left.\diff{u}{l}\right|_{l=1} &= (z+2\chi-2)u.\label{084845_20Jul24}
\end{align}
The diffusion parallel to the flow direction $D_\parallel$ does not
appear since the spatial derivative along that direction was already
dropped in the approximation~(\ref{204823_5Aug24}).

\subsection{Gaussian fixed point with shear}
As in the case of Model A, we require the scale invariance of
$\dot{\gamma}$, $D_\perp$, and $\Delta$:
\begin{align}
\left.\diff{\dot{\gamma}}{l}\right|_{l=1}
=\left.\diff{D_{\perp}}{l}\right|_{l=1}
=\left.\diff{\Delta}{l}\right|_{l=1}=0,
\end{align}
leading to 
\begin{align}
z+1-\zeta = z-4 = z-2\chi-(d-1+\zeta)-2 = 0.
\end{align}
Solving the above scaling relations, we get the following critical
exponents:
\begin{align}
\zeta = 5,\quad z = 4,\quad \chi = -\frac{d+2}{2}.
\end{align}
The anisotropic exponent satisfies $\zeta>1$, which justifies the
approximation Eq.~(\ref{204823_5Aug24}). The RG flow equation for the
non-linear term (\ref{084845_20Jul24}) is
\begin{align}
\left.\diff{u}{l}\right|_{l=1} = -du,
\end{align}
leading to $u\sim l^{-d}$. For $d>0$, $u\to 0$ in the thermodynamic
limit $l\to\infty$, meaning that the upper critical dimension is
\begin{align}
d_{\rm up}= 0.
\end{align} 
Therefore, the mean-field critical exponents are observed for any $d>0$.
Furthermore, the scaling exponent of the order parameter $\chi$ becomes
negative for $d>d_{\rm low}$ with
\begin{align}
d_{\rm low}= -2.
\end{align}
In particular, the continuous symmetry breaking can occur in $d=2$, as
in the case of Model A with shear.

% For $d>d_{\rm up}$, $u$ is a
% dangerous irrelevant variable.
% As in the case of Model A,
% a careful treatment of this variable leads to 
% \begin{align}
% \phi_a(\varepsilon,u)\sim \abs{\varepsilon}^{\beta},\quad \beta = \frac{1}{2}.
% \end{align}

\subsection{Correlation function}
The scaling behaviors, 
$\phi\sim l^{\chi}$, $x_1\sim l^\zeta$, $\bx_\perp\sim l$,
and $\varepsilon\sim l^{-(z-2)}$, 
lead to
\begin{align}
C(x_1,\bx_\perp,\varepsilon) &= \ave{\vec{\phi}(x_1,\bx_\perp)\cdot\vec{\phi}(0,0)}\new 
&=l^{2\chi}C(l^{-\zeta}x_1,l^{-1}\bx_\perp, l^{z-2} \varepsilon).
\end{align}
In the Fourier space, we get 
\begin{align}
C(q_1,\bm{q}_\perp,\varepsilon) &=
 \int d\bx e^{i(q_1x_1+\bq_\perp\cdot\bx_\perp)}C(x_1,\bx_\perp,\varepsilon)\new 
& =l^{2\chi+\zeta+d-1}
C(l^\zeta q_1,l\bm{q}_\perp,l^{z-2}\varepsilon)\new 
&=  l^{z-2}
C(l^\zeta q_1,l\bm{q}_\perp,l^{z-2}\varepsilon).\label{181236_20Jul24}
\end{align}
Substituting $l=\varepsilon^{-1/(z-2)}$, we get
\begin{align}
C(q_1,\bq_\perp,\varepsilon) =
 \varepsilon^{-1}C(\varepsilon^{-\nu_\parallel}q_1,\varepsilon^{-\nu_\perp}\bq_\perp,1)
\end{align}
with 
\begin{align}
\nu_\parallel = \frac{5}{2},\quad 
\nu_\perp = \frac{1}{2}.
\end{align}
The exponents are consistent with the previous mean-field analysis for
the model-H in
OK79~\cite{onuki1979,onuki1980,onuki1980-2,onuki2002phase}.  This is a
reasonable result, because the order parameter and hydrodynamic momentum
of the model-H decouple at the level of the linear analysis, leading to
the same equation of motion as that of Model B~\cite{onuki1979}. Also,
from Eq.~(\ref{181236_20Jul24}), we get
\begin{align}
&C(q_1,\bm{0},0) \sim \abs{q_1}^{-2/5},&\abs{q_1}\ll 1,\label{cb}\\
&C(0,\bq_\perp,0) \sim \abs{\bq_\perp}^{-2},&\abs{\bq_\perp}\ll 1,\\
&C(0,0,\varepsilon) \sim \varepsilon^{-1},&\varepsilon\ll 1,
\end{align}
implying the following expansion:
\begin{align}
C(q_1,\bq_\perp,\varepsilon) =
\left[c_1\varepsilon + c_2 \abs{q_1}^{2/5}
 + c_3 \abs{\bq_\perp}^2 + \cdots\right]^{-1},\label{182803_20Jul24}
\end{align}
which are again consistent with the theoretical prediction in OK79~\cite{onuki1979}.
The correction along the flow direction $C(q_1,\bm{0},0)\sim \abs{q_1}^{-2/5}$, Eq.~(\ref{cb}), is much smaller than that for Model A $C(q_1,\bm{0},0)\sim \abs{q_1}^{-2/3}$, Eq~(\ref{ca}), meaning that the critical fluctuations are more strongly suppressed than those of Model A.
This observation supports that Model B has smaller critical dimensions, $d_{\rm low}=-2$ and $d_{\rm up}=0$, compared to Model A, with $d_{\rm low}=0$ and $d_{\rm up}=2$.

\section{Summary and discussions}
\label{203110_28Jul24}

\begin{table}[t]
\begin{center}
\caption{Lower and upper critical dimensions, and critical exponents in
equilibrium and shear flow.  Note that the shear flow can be defined
only in $d\geq 2$.}  \label{144026_7Aug24}
\begin{tabular}{l|rr|rr}
&in equilibrium &  &  in shear flow &  \\ 
& Model A& Model B& Model A&Model B\\ \hline
$d_{\rm low}$ &2& 2&  0& -2 \\
$d_{\rm up}$ &4& 4&  2& 0\\
$\zeta$ & 1& 1&  3&5\\
$z$ & 2 & 4 &  2 &4 \\
$\chi$ &(2-d)/2& (2-d)/2& -d/2 & -(d+2)/2\\
$\nu_\parallel$ &1/2& 1/2& 3/2& 5/2\\
$\nu_\perp$ &1/2& 1/2& 1/2 & 1/2
\end{tabular}
\end{center}
\end{table}

In this work, we investigated the $O(n)$ model subjected to the steady
shear flow for both non-conserved (Model A) and conserved (Model B)
order parameters.  Using the dynamical RG analysis incorporating the
anisotropic scaling, we identify a new Gaussian fixed point that is
stable under the shear flow.  Table~\ref{144026_7Aug24} summarizes the
critical dimensions and exponents corresponding to this fixed point.  In
addition, we calculated the scaling behaviors of the correlation
functions near this Gaussian fixed point.  The correction in the Fourier
space behaves as $C(q_1)\sim \abs{q_1}^{-a}$, where $q_1$ denotes the
wave vector along the flow direction.  The exponent $a$ is smaller than
its equilibrium value of $2$ ($a = 2/3$ for Model A and $a = 2/5$ for
Model B), indicating that the shear flow suppresses the critical
fluctuations along the flow direction.  This suppression is consistent
with the reduction of the critical dimensions, $d_{\rm low}$ and $d_{\rm
up}$.

We found that $d_{\rm up}=2$ for Model A
and $d_{\rm up}=0$ for Model B~\footnote{Note however that the shear
flow Eq.~(\ref{024236_31Dec23}) can not be defined in $d<2$.}.  This
means that for both Model A and B, the critical exponents agree with
those of the mean-field predictions in $d=2$ and $3$.  As mentioned in
the Introduction, previous numerical studies in two-dimensional
non-conserved Ising model have often reported values of $\beta$ that
deviate from the mean-field critical exponent
$1/2$~\cite{cirillo2005,saracco2009,hucht2009,winter2010,angst2012}.
This discrepancy may arise from the presence of logarithmic corrections
at $d_{\rm up}$~\cite{nishimori2011elements}.  For the
conservative dynamics, a recent numerical result for the Ising model
shows a clear deviation from the mean-field
prediction~\cite{saracco2021critical}.  We speculate that this
discrepancy comes from the strong finite-size effect caused by the large
anisotropy exponent $\zeta=5$.  The large anisotropy exponent,
$\zeta=5$, implies that the system size needs to be scaled
anisotropically as $(L_\parallel, L_\perp) = (l^5, l)$ to allow for
sufficient development of critical fluctuations, where $L_\parallel$ and
$L_\perp$ are the linear sizes of the system parallel and perpendicular
to the flow direction, respectively.  Therefore, if we consider a system
with $L_\perp \approx 10$, $L_\parallel \approx 10^5$ is required to
accurately estimate the critical exponents.  The standard finite-size
scaling analysis, as used in Ref.~\cite{saracco2021critical}, would fail
to capture such extreme anisotropy.  It is interesting future work to
develop methods to precisely and efficiently calculate the critical
exponents of those systems by using numerical simulations.

Finally, we comment on the relation between the present analysis and the
earlier work by Onuki and Kawasaki (OK79)~\cite{onuki1979}. In OK79, the
authors studied Model~H under shear flow by assuming isotropic scaling,
while introducing anisotropy through wave-number--dependent renormalized
transport coefficients. Their renormalization-group equations
[Eqs.~(4.24)--(4.25) in OK79] depend on both amplitude and angle of wave
vector, and the resulting fixed point corresponds to the infinite
shear-rate limit. In contrast, the present study focuses on Models~A
and~B using an explicitly anisotropic-scaling framework. Because the two
analyses are based on different theoretical formalisms, their scaling
predictions are not directly comparable. A promising direction for
future work would be to apply OK79's approach to Models A and B, or
conversely, to extend the present anisotropic-scaling analysis to
Model~H. Such an investigation may help establish a matching procedure
between these complementary frameworks in a regime where both
descriptions are valid.

%  A promising direction for
% future work would be to apply OK79's approach to Models~A and~B, or
% conversely, to extend the present anisotropic-scaling analysis to
% Model~H, which could help bridge these complementary frameworks.

\section*{Acknowledgements}
We thank H.~Tasaki, Y.~Kuroda, M.~Hongo, and S.~Sasa for useful
discussions.  The authors thank YITP at Kyoto University and RIKEN
iTHEMS. Discussions during the workshop (YITP-T-24-04) on “Advances in
Fluctuating Hydrodynamics: Bridging the Micro and Macro Scales” were
useful in completing this work.

% % TODO: include author contributions
% \paragraph{Author contributions}
% This is optional. If desired, contributions should be succinctly described in a single short paragraph, using author initials.

% TODO: include funding information
\paragraph{Funding information}
This project has received JSPS KAKENHI Grant Numbers 23K13031.

\bibliography{reference.bib}

\nolinenumbers

\end{document}